\documentclass{PoS}
\title{Observational constraints on the X--ray Bright supergiant B[e] stars LHA 115-S18 \& LHA 120-S 134}

\ShortTitle{X--ray Bright sgB[e] stars}

\author{\speaker{Elizabeth S. Bartlett}\thanks{ESB is supported by a Claude-Leon Foundation Fellowship.}\\
        Astrophysics, Cosmology and Gravity Centre, Department of Astronomy, University of Cape Town, Rondebosch 7701, South Africa\\\
        E-mail: \email{ebartlett@ast.uct.ac.za}}

\author{J. Simon Clark\\
	Department of Physics and Astronomy, The Open University, Walton Hall, Milton Keynes, MK7 6AA, United Kingdom\\
        E-mail: \email{simon.clark@open.ac.uk}}

\abstract{We present the preliminary results of an ongoing series of spectroscopic observations of the Small Magellanic Cloud star LHA 115-S 18 (S18), which has demonstrated extreme photospheric and spectroscopic variability that, in some respects, is reminiscent of Luminous Blue Variables (LBVs). In contrast to our previously published results, between 2012--2015 S18 remained in an spectral state intermediate between S18's ``hot'' and ``cool'' extremes. In conjunction with contemporaneous OGLE--IV photometric monitoring of S18, these data will be used to determine the characteristic timescale of the variability and search for periodicities, in particular binary modulated periodicity. We also present the results of a pilot study of the LMC star LHA 120-S 134.}

\FullConference{SALT Science Conference 2015 -SSC2015-\\
		1-5 June, 2015\\
		Stellenbosch Institute of Advanced Study, South Africa}

\begin{document}

\section{Introduction}

\begin{figure}[t]\centering
\includegraphics[width=0.65\textwidth]{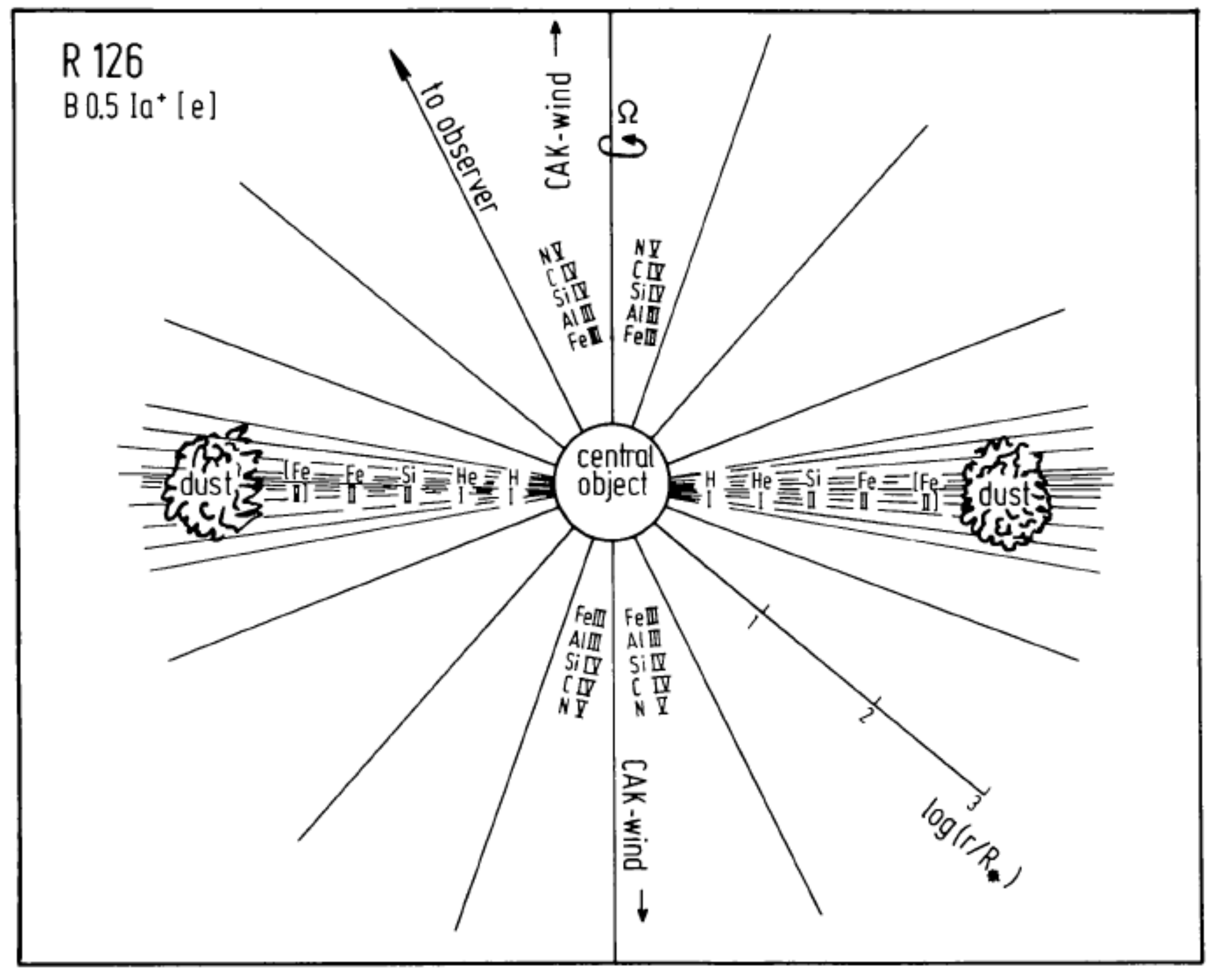}
\caption{Schematic of a sgB[e] star taken from Zickgraf et al. (1985) \cite{Zickgraf1985}}
\end{figure}\label{fig:b[e]}LHA 115-S 18 (S18) is a supergiant (sg) B[e] star in the Small Magellanic Cloud (SMC). Properties of stars exhibiting the B[e] phenomenon include strong Balmer emission lines, the presence of forbidden emission lines in their optical spectra and a large infrared (IR) excess. These are attributed to an extended circumstellar torus, with the IR excess originating from hot dust. Zickgraf et al. (1985, 1986) \cite{Zickgraf1985,Zickgraf1986} propose a two component stellar wind model to explain the optical spectra of sgB[e] stars in the Magellanic Clouds. The canonical model consists of a dense equatorial wind or torus exhibiting a low outflow velocity and a radiation driven wind in the polar region, indistinguishable from that of a normal supergiant (see Figure \ref{fig:b[e]}). We note here that, whilst the description above sounds similar to that of normal Be stars, the circumstellar environment of sgB[e] stars is dense enough to permit the formation/survival of dust. X--ray observations indicate that S18 is 'X--ray bright', i.e. its X--ray luminosity is an order of magnitude greater than expected for a massive single star of its bolometric luminosity (Antoniou et al. 2009 \cite{Antoniou2009}; Clark et al. 2013 \cite{Clark2013a}; Bartlett et al. in prep.). Binarity is typically invoked to explain such behaviour, with the excess high energy emission either arising via shocks in a colliding wind binary, in which the secondary is a massive star, or via accretion onto a compact companion. 

In recent years it has become apparent that the majority (\textgreater70\%) of the most massive stars do not exist in isolation but form and evolve in binary systems (Sana et al. 2012 \cite{Sana2012}). The interactions between the two companions have a profound affect on their evolution and, in turn, have important implications for other areas of astrophysics. There's mounting support for the idea that B[e] stars represent either interacting or newly formed, post--interaction binary systems (Kastner et al. 2010 \cite{Kastner2010}). Identifying sgB[e] binaries is, therefore, a priority as their properties will help constrain the physics of mass transfer.

\begin{figure}[t]\centering
\includegraphics[width=1.0\textwidth]{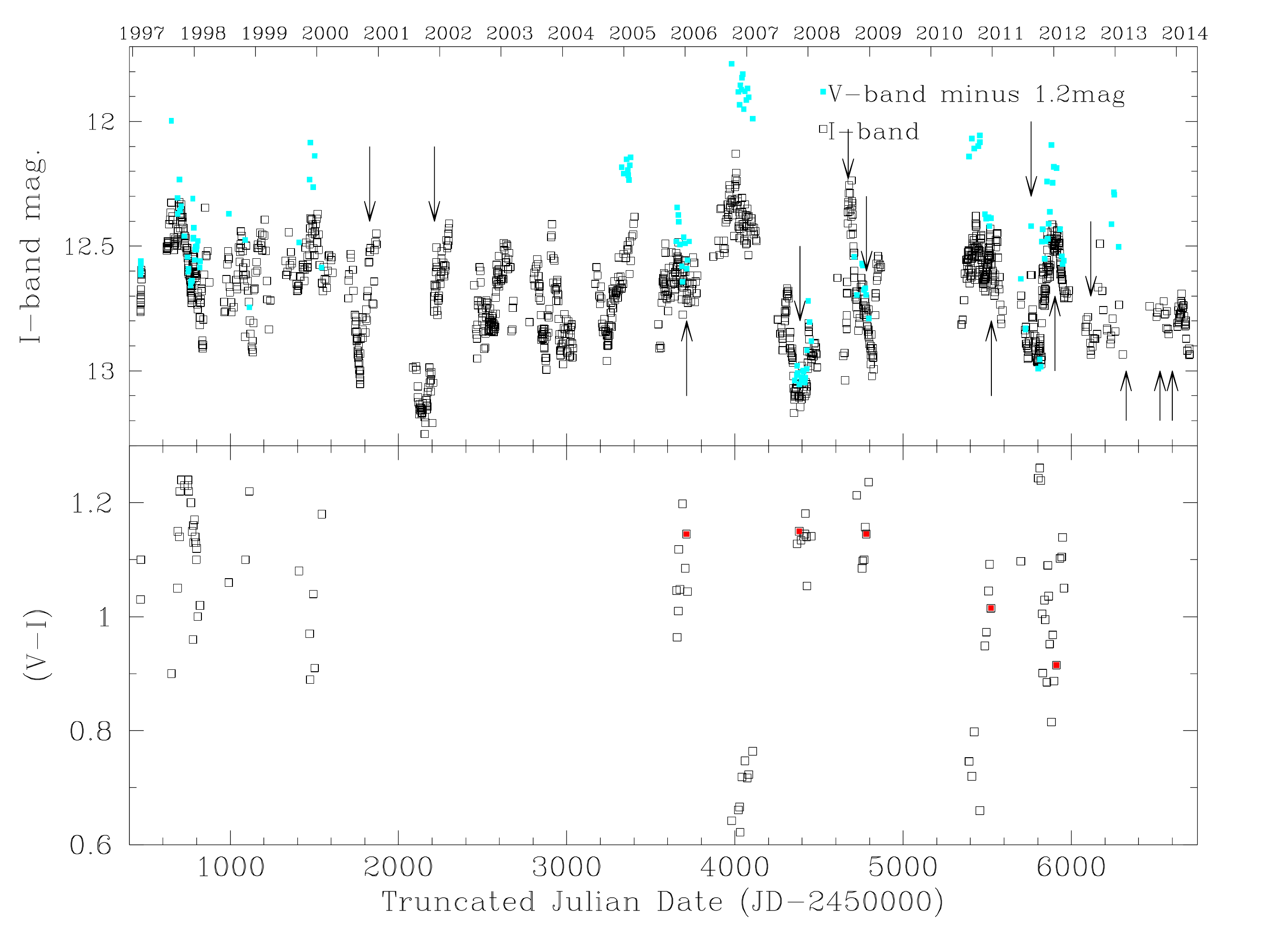}
\caption{Top panel shows the $V$ and $I$ band OGLE light curves. The $V$ band data are displaced by 1.2 mag to facilitate comparison. Bottom panel shows the ($V-I$) colour index against time. This plot is an updated version of Figure 3 in Clark et al. (2013) \cite{Clark2013a}.}\label{fig:ogle}
\end{figure} There are very few known sgB[e] stars with X--ray emission: XTE~J0421+560/CI Camelopard\-alis (CI Cam; e.g. Bartlett et al. 2013 \cite{Bartlett2013}), IGR~J16318-4848 (e.g. Barrag{\'a}n et al. 2009 \cite{Barragan2009}) and Wd1-9 (Clark et al. 2013 \cite{Clark2013b}). The first two sources are \emph{bona fide} High-Mass X--ray Binary (HMXB) systems, the third source a colliding wind binary. The HMXB GX301-2 (e.g. F{\"u}rst et al. 2011 \cite{Furst2011}), also gets an `honourary mention' in the first category -- whilst the optical counterpart is formally that of an early B-hypergiant and not a sgB[e] star (e.g. Clark et al. 2012 \cite{Clark2012}), this system shares many characteristics with CI Cam and IGR J16318-4848, including an infrared excess attributed to circumstellar dust, a strong 6.4~keV Fe--K$\alpha$ line and high absorbing column (n$_H\sim10^{23}$~cm$^{-2}$).

With the exception of Wd1-9, which resides in a well studied cluster, the distances to the Galactic X--ray emitting sgB[e] stars are plagued with uncertainties, which in turn leads to uncertainties in the source luminosity and thus the nature of the compact object. GX301-2 has clear pulsations detected at $\sim685$~s, unequivocally indicating that the accretor is a neutron star, but no spin periods have been detected thus far in CI Cam or IGR~J16318-4848. In contrast, the distance to the Magellanic Clouds are well constrained; any uncertainty as to the position of a source within these galaxies is trivial in comparison with their distance. As such, the population of sgB[e] stars in the Magellanic Clouds may be key to understanding these systems. 

Clark et al. (2013) \cite{Clark2013a} present an optical study of S18 based on a combination of `new' and archival spectroscopic data and the 16 year OGLE II--IV photometric curve (see Figure \ref{fig:ogle}). The authors show that, unlike normal sgB[e] stars which historically have been considered essentially static, S18 is highly variable. Both photometric ($\Delta$m$\sim1.3$mag) and spectroscopic variability (evolution from a P Cygni to WN9h spectral types implying $T_* \sim 19 \rightarrow 28 $kK) occurs with unprecedented rapidity; e.g. the  $\Delta$m$\sim 0.9$mag brightening in only $\sim$50~days. Moreover the spectral and photometric changes appear completely uncorrelated. 

Only one other sgB[e] star in the Magellanic Clouds is detected in X--rays: The Large Magellanic Cloud (LMC) star LHA 120-S 134 (S134, see Massey et al. 2014 \cite{Massey2014}). Interestingly it is also the only other known B[e] star Galaxy, SMC and LMC to show the broad feature around 4686\AA{} (Zickgraf et al. 1986 \cite{Zickgraf1986}).

To date observations of S18 been obtained on an ad hoc basis, resulting in a sparsely sampled, heterogeneous, low resolution data set. No modern multi epoch study of S134 has been made. Here we report our preliminary results of an ongoing Priority 4 SALT RSS monitoring campaign of S18, as well as a pilot study of S134, initiated in Semester 1 of 2014.

\section{S18}
\begin{figure}[t]\centering
\includegraphics[width=1.0\textwidth]{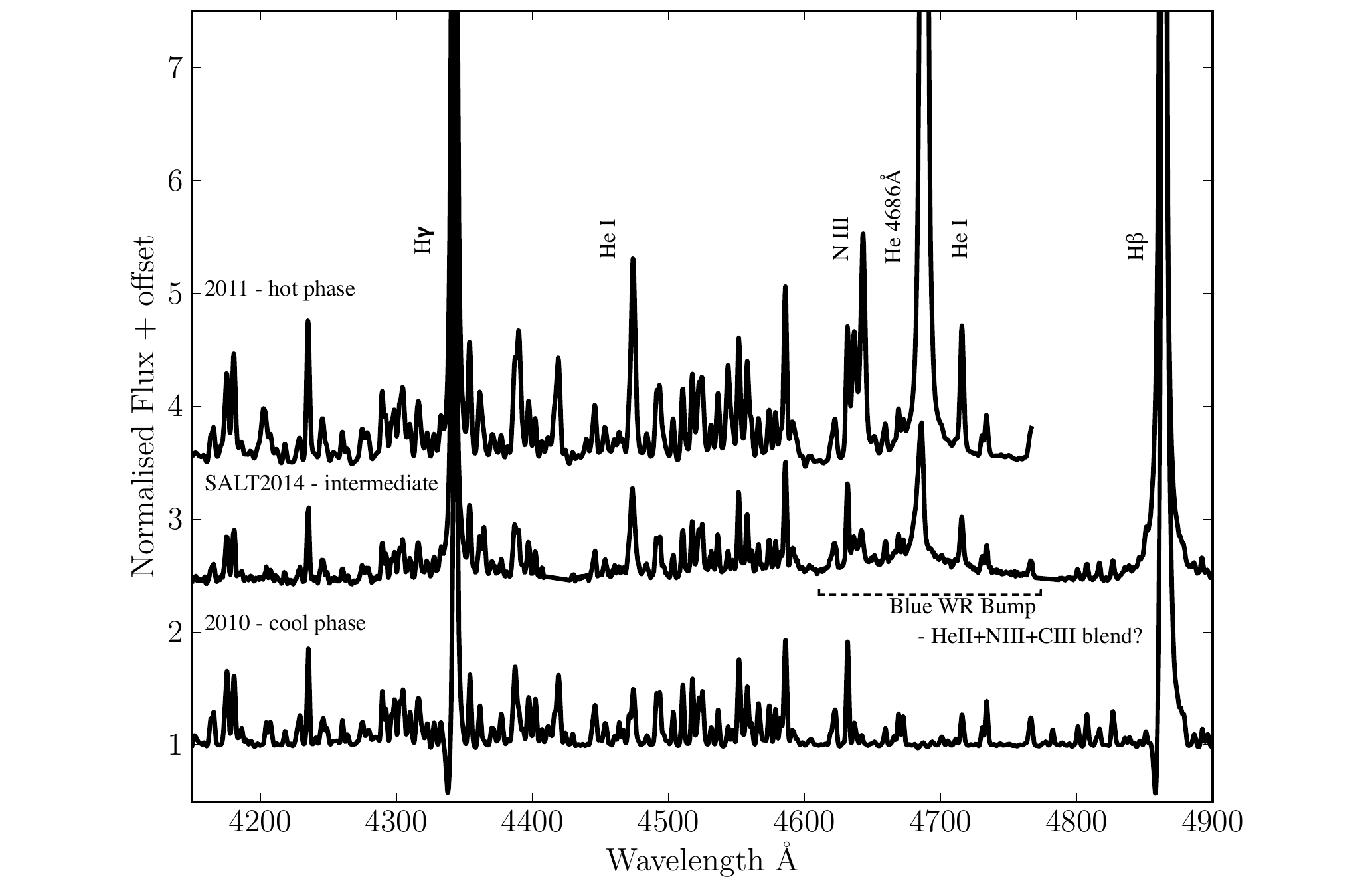}
\caption{Spectra of S18 around the He\textsc{ii} in its cool state (2010), hot state (2011) and the apparent intermediate state seen in SALT spectra (taken 2014 July 22). The main spectral transitions are labelled.  Whilst the morphology of S18 is similar to that of a late type, Nitrogen-rich Wolf-Rayet (WNLh) star, the line emission is far stronger than any known example.}\label{fig:multi}
\end{figure} Figure \ref{fig:multi} shows a representative SALT spectrum from our monitoring campaign, along with archival spectra, taken in 2010 and 2011. In our previous data S18 was typically observed in either a ``hot'' or ``cool'' extreme, but in our observations between 2014--2015 we found S18 occupied neither state, but rather demonstrated a spectral morphology intermediate between the two. Note the evolution in the Balmer series from a P Cygni profile to pure, stronger emission and the appearance of N\textsc{iii} and He\textsc{ii}. Whilst the morphology of S18 is similar to that of a WNLh star, the line emission is far stronger than any known example.

The data, along with SALT observations of a comprehensive sample of WN Wolf--Rayets taken for comparison, indicate that the strength and the line widths in the high excitation lines in S18 (He\textsc{ii} \& N\textsc{iii}) are inconsistent with a simple blend of B supergiant  and N--rich WR. Instead the exciting radiation from the central engine (whether accretion or colliding wind binary) must ionise the circumstellar torus and associated disc wind that it drives, yielding the composite emergent spectrum.

\begin{figure}[t]
\centering
\includegraphics[width=0.75\textwidth]{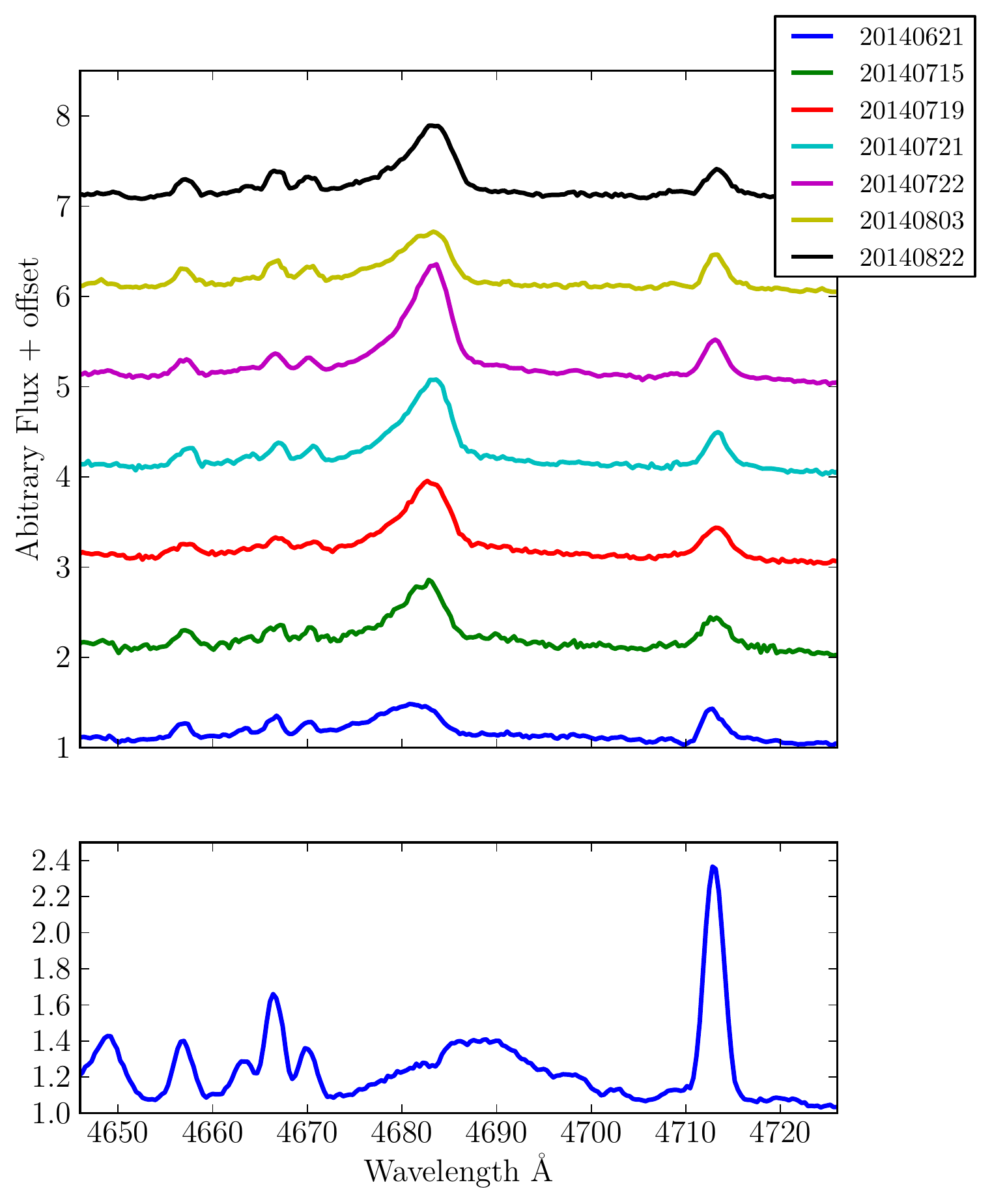}
\caption{SALT spectra of S18 (top panel) and S134 (bottom panel) around the He\textsc{ii} 4686\AA{} region}\label{fig:SALT}
\end{figure}

The top panel of Figure \ref{fig:SALT} shows the He\textsc{ii} 4686\AA{} region of the SALT spectra of S18, taken in Semester 1 of 2014. The data reveal unprecedented variability, on day to day timescales, compared with months to years from our previous monitoring.  We also note the transient P Cygni profile in the He\textsc{i}  series e.g. 4710\AA{} line.

\subsection{S18 as a Luminous Blue Variable/Supernova Imposter}

Luminous Blue Variables (LBVs) are thought to represent a short-lived phase of very massive stellar evolution in which instabilities drive dramatic mass loss. They occupy the same region of the H-R diagram as sgB[e] stars and appear to be the immediate progenitors of some of the most luminous SNe known (e.g. SNe 2005gl and 2006tf; Gal-Yam et al. 2009 \cite{Gal-Yam2009}; Smith et al. 2008 \cite{Smith2008}). By virtue of its dramatic spectroscopic and photometric variability, S18 trivially satisfies the eponymous classification criteria of LBVs, but does not conform to their typical behaviour (Humphreys \& Davidson, 1994 \cite{Humphreys1994}) and more closely resembles the SNe impostors SN2000ch and SN2009ip. 

The LBV/colliding wind binary HD~5980 -- an obvious comparator to S18 -- shows  modulation of its very strong He\textsc{ii} 4686\AA{} line phased to the 19.3 day orbital period (e.g. Koenigsberger et al. 2010 \cite{Koenigsberger2010}). Based on the short term variability in the He\textsc{ii} line in S18 revealed by SALT, one might ask whether periodic modulation is present on a similarly short period and hence whether S18 is also a compact colliding wind binary.

He\textsc{ii} 4686\AA{} has yet to reach its historical maximum intensity during our SALT observations. Intriguingly, $\eta$ Carinae is a binary with a period of 5.54 years that, again, is delineated by variability in the He\textsc{ii} 4686\AA{} emission (Teodoro et al. 2012 \cite{Teodoro2012}). Could comparable long term behaviour also be present in S18, superimposed on the rapid line profile variability? Extending this idea, the SNe impostors that S18 resembles by virtue of its rapid photometric variability (e.g. SN2000ch and SN2009ip) are both He\textsc{ii} 4686\AA{} emitters that show dramatic outbursts separated by years of quiescence. 

\section{S134}

The bottom panel of Figure \ref{fig:SALT} shows a representative SALT spectrum of S134, taken in Semester 2 of 2014. Unlike S18,  the overall strength of the He\textsc{ii} 4686\AA{}  remained roughly constant over the $\sim3$~month period of the observations, although we note that the line profile is variable. However, we recognise that several studies of S18 taken in isolation would also be consistent with a lack of variability in this line.

We know far less about S134 than S18 due to the lack of long term monitoring, although [Ne\textsc{iii}], [S\textsc{iii}] and [S\textsc{iv}] emission in the mid-IR Kastner et al. (2010) \cite{Kastner2010} imply excitation temperatures of \textgreater60kK; in conjunction with both X--ray and  He\textsc{ii} 4686\AA{} emission providing a compelling case for binarity (Bartlett et al. in prep.). Our  SALT observations of S134 and benchmarking WR stars show that the spectrum of S134 is consistent with a composite blend providing a compelling case for binarity as suggested by Massey et al. (2014) \cite{Massey2014}. Our SALT observations of S134 and benchmarking WR stars show that the spectrum of S134 is consistent with a composite blend (Bartlett et al. in prep.)

\section{Conclusions}

Despite having many similarities (i.e. X--ray emission and broad emission feature at 4686\AA{}) our SALT observations of S18 and S134 thus far suggest that they are very different binary systems. The He\textsc{ii} 4686\AA{} line in the spectrum of S134 is consistent with us directly observing a WNL star in the system, whereas the strength of this line in the spectrum of S18 precludes this. Along with its rapid variability, the strength of this line suggest S18 has more in common with the SNe impostors SN2000ch and SN2009ip. At the time of writing, we have neither the cadence nor sufficient observations to search for short or long term periodic modulation with our current data set to link S18 with such systems.

Data obtained since this meeting indicates that S18 currently completely lacks He\textsc{ii} 4686\AA{} emission; a state only observed four times in the past 60 years. The inferred absence of high energy exciting radiation means that we will be able to determine the quiescent outflow velocity and mass loss rate from S18 via modelling of the  P Cygni profiles of the Balmer lines, providing a benchmark to investigate  the effect of ionisation by the central (binary star?) engine as it turns back on again. 

More generally, this campaign highlights the power of SALT Priority 4 programs, that is, programs undertaken in more marginal conditions that do not strictly require SALT's aperture. Even during P4 intervals, SALT can provide homogeneous spectra of unprecedented signal to noise and resolution. During these observations we also serendipitously identified a rare LBV eruption in our WN10h star (HDE 269582 == BAT99-45 == LHA 120-S 83), highlighting the utility of our P4 proposal.

\bibliographystyle{JHEP}
\bibliography{SSC2015_055}

\end{document}